\documentclass[12pt]{iopart}
\usepackage{iopams} 

\usepackage{times}

\DeclareFontFamily{OT1}{times}{}
\DeclareFontShape {OT1}{times}{m }{n }{ <-> ptmr }{}
\DeclareFontShape {OT1}{times}{bx}{n }{ <-> ptmb }{}
\DeclareFontShape {OT1}{times}{m }{it}{ <-> ptmri}{}
\DeclareFontShape {OT1}{times}{bx}{it}{ <-> ptmbi}{}

\newcommand{\DEF}{:=}   
\newcommand{\eqref}{\eref}
\newcommand{\DUP}{\delta}             
\newcommand{\UPS}{\Upsilon}           
\newcommand{\BRA}{\langle\kern -.2em\langle} 
\newcommand{\KET}{\rangle\kern -.2em\rangle} 
\newcommand{ \iint}{\int\kern -.8em\int}
\newcommand{\iiint}{\int\kern -.8em\int\kern -.8em\int}
\newcommand{\dfrac}{\frac}
\newcommand{\sgn}{{\rm sgn}} 
\newcommand{\notag}{\nonumber}
\renewcommand{\theequation}{\thesection.\arabic{equation}}

\begin{document}

\title[On the electromagnetic momentum of static charge
       and  steady current  distributions]
      {On the electromagnetic momentum of static
      charge and  steady current  distributions}

\author{Andre Gsponer}

\address{Independent Scientific Research Institute,
         Oxford, OX4 4YS, United Kingdom}

\begin{abstract}

Faraday's and Furry's formulae for the electromagnetic momentum of static charge  distributions combined with steady electric current distributions are generalised in order to obtain full agreement with Poynting's formula in the case where all fields are of class $\mathcal{C}^1$, i.e., continuous and continuously differentiable, and the integration volume is of finite or infinite extent.

These three formulae are further generalised to the case where singularities are allowed to exist at isolated points in the fields, and at surfaces separating domains in which the distributions are of class $\mathcal{C}^1$.

Applications are made to electric and magnetic, point-like and finite dipolar systems, with an emphasis on the impact of singularities on the magnitude and location of the electromagnetic momentum.

\end{abstract}

\noindent{\bf Published:} Eur. J. Phys. {\bf 28} (2007) 1021--1042.

\section{Introduction}
\label{int:0} \setcounter{equation}{0}

As recently recalled by Jackson \cite{JACKS2006-} and McDonald \cite{MCDON2006-}, for systems in which effects of radiation and of retardation can be ignored, the total electromagnetic momentum can be calculated in various equivalent ways,
\begin{eqnarray}
\label{int:1}
    \vec{P}_{\rm Poynting}
      &= \iiint d^3V \dfrac{\vec{E} \times \vec{B}}{4 \pi},\\
\label{int:2}
    \vec{P}_{\rm Furry}
      &= \iiint d^3V \dfrac{\phi \vec{j}}{c^2},\\
\label{int:3}
    \vec{P}_{\rm Faraday}
      &= \iiint d^3V \dfrac{\rho \vec{A}}{c},
\end{eqnarray}
where $\rho$ is the electric charge density, $\vec{A}$ is the magnetic vector potential (in the Coulomb gauge where $\vec{\nabla} \cdot \vec{A} = 0$), $\vec{E}$ is the electric field, $\vec{B}$ is the magnetic field strength, $\phi$ is the electric (scalar) potential, $\vec{j}$ is the  electric current density, and $c$ is the speed of light (which we take as $c=1$ in the following).  Finally, it is generally assumed that these formulae apply when the integrals extend over the whole three-dimensional space, and when all the integrands decrease sufficiently fast at infinity.

   In this paper we show that the electromagnetic momentum of a finite or point dipole in an electromagnetic field can be consistently calculated with the three equivalent formulae (\ref{int:1}--\ref{int:3}), as well with their more general forms applicable when the integration volume is finite, using the methods presented in reference \cite{GSPON2004D}.\footnote{There are a few misprints in that paper which are corrected in the errata given at the end of the present paper.}  These methods allow a straightforward calculation in all cases, including those in which there are singularities in the fields.

The interest of calculating the electromagnetic momentum of a stationary dipole in a constant electromagnetic field is not just to confirm that such systems have a ``hidden'' electromagnetic momentum even though they are static charge distributions combined with steady electric currents \cite{FURRY1969-, HNIZD1997-}:  these investigations confirm that the electromagnetic fields of point singularities have to be supplemented by $\DUP$-like contributions which cannot be neglected in the general case \cite{TANGH1962-,JACKS1977-,BARUT1993-,GSPON2006C}.

For example, the complete field strength of a point magnetic dipole of magnetic moment $\vec{m}$ is \cite{GSPON2004D} 
\begin{eqnarray}\label{int:4}
    \vec{B}_{d}(\vec{r}) =  
        \Bigl( 3\frac{\vec{u}(\vec{m} \cdot \vec{u})}{r^3}
              - \frac{\vec{m}}{r^3} \Bigr)  \UPS(r)
              + \frac{\vec{u} \times 
                     (\vec{m} \times \vec{u})}{r^2}\DUP(r),
\end{eqnarray}
where $\vec{u}$ is the unit vector in the direction of $\vec{r}$, and, as will be recalled in section \ref{inc:0}, the generalised function $\UPS$ insures that differentiation at the position of the singularities properly leads to the $\DUP$ functions which arise when calculating the fields and currents according to Maxwell's equations.  For instance, after volume integration, the $\DUP$ singularity in \eqref{int:4} gives the $\DUP$-like term discussed  by Jackson in \cite[p.184]{JACKS1975-}, which is essential in calculating the hyperfine splitting of atomic states \cite{JACKS1977-}.

Similarly, the complete field of a point electric dipole (which is also called electric `dimonopole' to stress that its distributional properties are radically different from those of an intrinsic magnetic `dipole') of moment $\vec{p}=q\vec{d}$, where $\vec{d}$ is the infinitesimal vector separating the positions of two poles of charge $\pm q$, is \cite{GSPON2004D}
\begin{eqnarray}\label{int:6}
    \vec{E}_{dm}(\vec{r}) = 
    \Bigl( 3\frac{\vec{u}(\vec{p} \cdot \vec{u})}{r^3}
          - \frac{\vec{p}}{r^3} \Bigr)  \UPS(r)
  -  \frac{\vec{u}(\vec{p}\cdot\vec{u})}{r^2} \DUP(r).
\end{eqnarray}
Here we have the $\DUP$ singularity that after volume integration yields the $\DUP$-like term discussed by Jackson in \cite[p.~141]{JACKS1975-}, which together with the corresponding one in \eqref{int:4} explicitly appears in the formulation of quantum electrodynamics of point particles having anomalous electromagnetic moments \cite{BARUT1993-}.  In section \ref{mag:0} it will be seen that this $\DUP$ singularity is also indispensable to calculate the electromagnetic momentum using (\ref{int:1}--\ref{int:3}) for a point dipole.

The plan of this paper is as follows:  

In section \ref{gen:0} we derive the Furry and Faraday formulae and keep them in the general form including the surface terms.  It is assumed that all potentials, fields, and source charge or current distributions are well behaved functions, i.e., differentiable with a continuously differentiable derivative at each point.

In section \ref{inc:0} we consider the case where the potentials, fields, and sources may have singularities, such as those at the locations of point charges or dipoles, as well as those at the boundaries of idealised  finite systems.

In section \ref{ele:0} we calculate the electromagnetic momentum of a point electric dipole in a locally uniform magnetic field $\vec{B}_0$.  It will be found that the dipole's momentum is $\vec{B}_0 \times \vec{p}/3$, provided the surface terms are included in the Faraday and Furry formulae, and provided the $\DUP$-function term appearing in the full expression of the electric dipolar field, i.e., equation \eqref{int:6}, is included when evaluating Poynting's expression.

In section \ref{mag:0} we calculate the electromagnetic momentum of a point magnetic dipole in the field of a point electric charge.  It will be found with all three methods that the total momentum is $\vec{E}_0 \times \vec{m}$, where $\vec{E}_0$ is the electric field at the position of the dipole, provided the  $\DUP$-function term appearing in  equation \eqref{int:4} is included when evaluating Poynting's expression.

In section \ref{fin:0} a finite electric and magnetic dipolar configuration is considered: two concentric shells, one ``dipole charged'' and the other ``dipole magnetised'' \cite{ROMER1995-}. Since such shells produce a field that is dipolar outside them and uniform inside, this configuration provides an idealised model for a finite electric dipole locate within a uniform magnetic field, or vice versa, which has the advantage to be exactly solvable \cite{HNIZD1997-}.  It will be found with Furry's and Faraday's methods that the total electromagnetic momentum is equal to $\vec{B}_0 \times \vec{p}/2$. This result will be confirmed with Poynting's method, which will also enable us to find out where the electromagnetic momentum is located, as well as the respective amounts of the various regular (i.e., field-like) and singular (i.e., particle-like) contributions to this momentum.

Finally, in section \ref{con:0}, we come back on our main results and end with a brief discussion of the broader implications of the results obtained by applying Poynting's method to the double-shell model of section \ref{fin:0}, which show that the integrated electromagnetic momentum can be considered as being entirely located in the fields, or entirely bound to the sources, or else partially located in the fields and partially bound to the sources.

\section{Generalised Furry and Faraday formulae}
\label{gen:0} \setcounter{equation}{0}

In this section we derive Furry's and Faraday's formulae in their general form starting from Poynting's formula \eqref{int:1}, i.e.,
\begin{eqnarray}
\label{gen:1}
    \vec{P}_{\rm Poynting}
      = \iiint d^3V \dfrac{\vec{E} \times \vec{B}}{4 \pi}.
\end{eqnarray}
This assumes that Poynting's formula is the fundamental expression of the electromagnetic momentum, which is generally accepted to be the case in the relativistic theory, and which is taken as the starting point in Jackson's derivation of these formulae \cite{JACKS2006-}.  However, as can be seen for example in Furry's paper \cite{FURRY1969-}, there is much flexibility to move between the three formulae, provided some supplementary assumptions are made.  It is in order to clarify these assumptions that we rederive Furry's and Faraday's formulae here.

Also, we would like to clarify to what extent the existence of singularities in the integration volume over which Poynting's formula is evaluated affects the validity of Furry's and Faraday's formulae.  In view of this we will first derive them assuming that there are no singularities, and consider the case where there are singularities in section \ref{inc:0}.

As will be seen, the main tool for deriving Furry's and Faraday's formulae from Poynting's (and vice versa) is Gauss's theorem.  In order to make integrations by parts and use that theorem to transform volume integrals into surface integrals it is necessary that the integrands should be sufficiently smooth functions, i.e., of class $\mathcal{C}^1$, that is differentiable with a continuously differentiable derivative at each point of the volume.  We will therefore agree that the symbols $V$ and $d^3V$ correspond to a simply connected volume, and that the integrands are finite and of class $\mathcal{C}^1$ at all points within it.

Concerning the validity of all three formulae there is also the question of the convergence of the integrals, especially when the integration volume is infinite.  As the answer to this depends on the specific integrand, we will for simplicity give only a sufficient condition, based on the demand that all functions are ``decreasing sufficiently fast at infinity.''  For instance, the criterion
\begin{eqnarray}
\label{gen:2}
    |\vec{E}|~|\vec{B}| (r \rightarrow \infty)
                = \mathcal{O} (r^{-3-\epsilon}),
\end{eqnarray}
where $0 < \epsilon < 1$, insures that the integral in Poynting's formula behaves as $\mathcal{O} (r^{-\epsilon})$ at infinity.  Similar criteria will be given for Furry's and Faraday's formulae to specify sufficient conditions under which the surface terms can be neglected.

\subsection{General Furry formula}

Let us express $\vec{E}$ in terms of the electric potential as $\vec{E} = -\vec{\nabla} \phi$.  This enables us to rewrite \eqref{gen:1} as
\begin{eqnarray}
\label{gen:3}
    \vec{P}_{\rm Poynting} = -\dfrac{1}{4\pi} \iiint d^3V
     \phi \overleftarrow{\nabla} \times \vec{B},
\end{eqnarray}
where the arrow indicates that the gradient operator acts towards the left.   To be able to make an integration by parts and use Gauss's theorem we need the gradient to operate on the whole integrand, that is on both sides of $\nabla$  in the last equation.  This is done by rewriting \eqref{gen:3} as
\begin{eqnarray}
\label{gen:4}
    \vec{P}_{\rm Poynting} = -\dfrac{1}{4\pi} \iiint d^3V
     \phi \overrightarrow{\overleftarrow{\nabla}} \times \vec{B}
                                 + \dfrac{1}{4\pi} \iiint d^3V
     \phi \overrightarrow{\nabla} \times \vec{B},
\end{eqnarray}
where the gradients operate as indicated.  We can now use Gauss's theorem to replace the first volume integral by a surface integral, and Maxwell's equation ${\nabla} \times \vec{B}  = 4\pi \vec{j}$, to get Furry's formula in its general form, i.e.,
\begin{eqnarray}
\label{gen:5}
    \vec{P}_{\rm Furry} = -\dfrac{1}{4\pi} \iint
     \phi d^2\vec{S} \times \vec{B}
                                 + \iiint d^3V
     \phi \vec{j}.
\end{eqnarray}
This expression is equal to Furry's simplified formula \eqref{int:2} when the surface term can be neglected, that is when
\begin{eqnarray}
\label{gen:6}
    |\phi~\vec{B}| (r \rightarrow \infty) = \mathcal{O} (r^{-2-\epsilon}),
\end{eqnarray}
because the surface element is $\mathcal{O} (r^2)$.

\subsection{General Faraday formula}

This time we express $\vec{B}$ in terms of the vector potential as $\vec{B} = \nabla \times \vec{A}$, and we use the identity 
\begin{eqnarray}
\label{gen:7}
         \vec{E} \times (\overrightarrow{\nabla} \times \vec{A})
    =  \underline{\overrightarrow{\nabla} (\vec{A}}\cdot\vec{E})
    -  (\vec{E}\cdot\overrightarrow{\nabla}) \vec{A},
\end{eqnarray}
where the underline indicates the range of the gradient operator.  This enable us to rewrite \eqref{gen:1} as
\begin{eqnarray}
\label{gen:8}
    \vec{P}_{\rm Poynting} = \dfrac{1}{4\pi} \iiint d^3V
     \underline{\overrightarrow{\nabla} (\vec{A}}\cdot\vec{E})
                               -\dfrac{1}{4\pi} \iiint d^3V
     (\vec{E}\cdot\overrightarrow{\nabla}) \vec{A}.
\end{eqnarray}
In the first integral we use $\vec{E} = -\vec{\nabla} \phi$, and we rewrite the second one in terms of a left-right operating gradient so we can later use Gauss's theorem. Thus 
\begin{eqnarray}
\label{gen:9}
    \vec{P}_{\rm Poynting} = &-\dfrac{1}{4\pi} \iiint d^3V
     \underline{\overrightarrow{\nabla}
            (\vec{A}}\cdot\overrightarrow{\nabla}\phi)\\
\label{gen:10}
                             &-\dfrac{1}{4\pi} \iiint d^3V
     (\vec{E}\cdot\overleftarrow{\overrightarrow{\nabla}}) \vec{A}
                                +\dfrac{1}{4\pi}  \iiint d^3V
     (\vec{E}\cdot\overleftarrow{\nabla}) \vec{A}.
\end{eqnarray}
For the same reason we introduce a left-right operating gradient in the first integral, i.e.,
\begin{eqnarray}
\label{gen:11}
    \vec{P}_{\rm Poynting} = &-\dfrac{1}{4\pi} \iiint d^3V
     \underline{\overrightarrow{\nabla}
          (\vec{A}}\cdot\overleftarrow{\overrightarrow{\nabla}}\phi)
                                 +\dfrac{1}{4\pi}  \iiint d^3V
     \underline{\overrightarrow{\nabla}
               (\vec{A}}\cdot\overleftarrow{\nabla}\phi)\\
\label{gen:12}
                             &-\dfrac{1}{4\pi} \iiint d^3V
     (\vec{E}\cdot\overleftarrow{\overrightarrow{\nabla}}) \vec{A}
                                +\dfrac{1}{4\pi}  \iiint d^3V
     (\vec{E}\cdot\overleftarrow{\nabla}) \vec{A}.
\end{eqnarray}
If we now postulate the Coulomb gauge, i.e.,
\begin{eqnarray}\label{gen:13}
        \nabla\cdot\vec{A}  = 0,
\end{eqnarray}
the integral on the right of \eqref{gen:11} is zero because partial differentiations commute.  As for the integral on the right of \eqref{gen:12} we use  $\vec{E}\cdot\overleftarrow{\nabla} = 4\pi\rho$.  It comes
\begin{eqnarray}
\label{gen:14}
    \vec{P}_{\rm Poynting} = &-\dfrac{1}{4\pi} \iiint d^3V
     \underline{\overrightarrow{\nabla}
          (\vec{A}}\cdot\overleftarrow{\overrightarrow{\nabla}}\phi)\\
\label{gen:15}
                             &-\dfrac{1}{4\pi} \iiint d^3V
     (\vec{E}\cdot\overleftarrow{\overrightarrow{\nabla}}) \vec{A}
                                +  \iiint d^3V \rho \vec{A},
\end{eqnarray}
where we just need to use Gauss's theorem twice to replace the first two volume integrals by surface integrals to get the general form of Faraday's formula, i.e.,
\begin{eqnarray}
\label{gen:16}
    \vec{P}_{\rm Faraday} = -\dfrac{1}{4\pi} \iint \Bigl[
 \phi\underline{\overrightarrow{\nabla} (\vec{A}}\cdot d^2\vec{S})
                  + (\vec{E}\cdot d^2\vec{S}) \vec{A} \Bigr]
                                +  \iiint d^3V \rho \vec{A}.
\end{eqnarray}
This expression is equal to Faraday's simplified expression \eqref{int:3} when both surface terms can be neglected, that is when
\begin{eqnarray}
\label{gen:17}
    |\phi~\partial_rA_r| (r \rightarrow \infty)
     = \mathcal{O} (r^{-2-\epsilon}),
    \quad
    |\vec{E}||\vec{A}| (r \rightarrow \infty)
     = \mathcal{O} (r^{-2-\epsilon}),
\end{eqnarray}
where $\partial_rA_r$ is the partial derivative of the radial component of $\vec{A}$.  As for the Coulomb gauge, equation \eqref{gen:13}, it is a necessary condition for the applicability of both the general \eqref{gen:16} and the simplified \eqref{int:3} forms of Faraday's formula.

\section{Inclusion of singularities}
\label{inc:0} \setcounter{equation}{0}

The simplified expressions (\ref{int:1}--\ref{int:3}) and the more general formulae \ref{gen:5} and \ref{gen:16} are rigorously applicable to any configuration of static charges and steady currents provided all potentials, field strengths, and source charge or current distributions are of class $\mathcal{C}^1$.  This means that they can be directly applied to any engineered electrical system in which all components have a geometry described by smooth finite functions.  They can also be applied to systems in which, for the purpose of simplifying the calculations, some components are idealised with Dirac delta-functions and Heaviside step-functions, in which case spurious singularities may arise as artifacts due to these idealisations, but can in general be discarded on the basis of simple plausibility arguments.

   On the other hand if these expressions are applied to systems including  elementary objects such as point-charges and point-dipoles there will be real singularities which in principle invalidate the application of these formulae since there will be points at which the integrands will no more be finite.  In such cases the general method consists of isolating the singularities in small regions enclosing them, and of dealing separately with the regular and singular parts of the integrals.  However, as is well known, the proper way of dealing with the singular parts is still an unsolved problem in electrodynamics, and the best that can be done is to use a prescription to ``regularise'' the integrals and possibly ``renormalise'' the results if so required.\footnote{This is obviously not very satisfactory, but at present the only way to justify the assumption that all distributions considered by electrical engineers are of class $\mathcal{C}^1$, which implies that the singular contributions stemming from the  charge and magnetic moment of all electrons are systematically ignored.} This is what we will do in sections \ref{ele:0} and \ref{mag:0}.

   There is another class of singularities which arise when an idealised configuration such as a charged-shell is considered as an elementary system, that is when it is considered as a genuine elementary solution of Maxwell's equation, even though it is not believed that such a configuration necessarily corresponds to a real physical object (in the sense that a point-charged is believed to correspond to an electron).  In this case, as will be seen in the example studied in section \ref{fin:0}, the integrals can be finite because the sources are no more concentrated at a point, but there may still be singularities at the boundaries (e.g., at the surface of a shell) which have to be physically interpreted.  Also, the integrals have to be evaluated separately in each domain defined by the boundaries.  For the Furry and Faraday formulae this implies that the surface terms have to be carefully evaluated at all boundaries to insure that their respective contributions are such that the total given by the outermost boundary is physically meaningful.

   Finally, when considering systems in which singularities overlap, or in which elementary objects self-interact, there will appear products of distributions which cannot be dealt with using standard distribution theory.  While new methods exists to evaluate these products \cite{GSPON2006B}, degenerate or self-interacting systems will not be considered in the present paper.

   Since the problems associated with singularities are complex and crucially depending on the details of the system under consideration, there is no general method to deal with them.  Nevertheless, some simplification is achieved by characterising the singularities in such a way that their origin is made explicit starting from the potentials of the fields.   This has the advantage that the singularities do not have to be discovered indirectly on a case-by-case basis by looking at the details of all expressions and their derivatives, but arise consistently and automatically in accord with the principles of the theory of the distributions. 

   A method to do that is explained in reference \cite{GSPON2004D}, which results in expressions such as equations \ref{int:4} and \ref{int:6} of the introduction.  The basis of this method is to make explicit the difference between the linear coordinate variables ($r$ in a polar system; $x$, $y$, and $z$ in a Cartesian system) which are continuous, and the modulus  $|\vec{r}|$ of the position vector ($|r|$ in a polar system; $\sqrt{x^2+y^2+z^2}$ in a Cartesian system) which is discontinuous at $\vec{r} = 0$.  In a polar coordinate system this is done by writing, for any occurrence of the radius vector $\vec{r}$,
\begin{eqnarray}\label{inc:1}
     \vec{r} = r \Upsilon(r) \vec{u}(\theta,\phi), 
\end{eqnarray}
where $\vec{u}$ is the unit vector in the direction of $\vec{r}$. Here, the generalised function $\UPS(r)$ has the properties
\begin{eqnarray}\label{inc:2}
    \UPS(x) \DEF 
         {
         \cases{
                \mbox{undefined} &for   $r < 0$,\\
                            0    &for   $r = 0$,\\
                           +1    &for   $r > 0$,
               }
         }
              \qquad\mbox{and}\qquad  \frac{d}{dx}\UPS(x) = \DUP(x),
\end{eqnarray}
as well as
\begin{eqnarray}\label{inc:3}
              \int_{0}^{\infty} dr~ \UPS(r) F(r)
          = \int_{0}^{\infty} dr~ F(r),
\end{eqnarray}
which is the counter-part of the defining property of the $\DUP$ function in polar coordinates, i.e.,\footnote[1]{We adhere to the modern convention that the symbol $\DUP$ corresponds to the \emph{Dirac measure,} i.e., that its defining property is $\BRA \delta | F \KET = F(0)$ independently of the support of $\DUP$, which is here the half-space $[0,\infty]$ rather than $[-\infty,\infty]$ as in the customary definition of Dirac's delta-function.}
\begin{eqnarray}\label{inc:4}
  \int_0^\infty dr~ \delta(r)F(r) = F(0),
\end{eqnarray}
where, like in \eqref{inc:3}, $F \in \mathcal{C}^1$ for the purpose of this paper.

\section{Electric dipole in a locally uniform magnetic field}
\label{ele:0} \setcounter{equation}{0}

In this section we consider an electric dipole located in a magnetic field that is uniform over a finite volume. That field may be generated by an external system such as a solenoid, or else the volume may be assumed to be sufficiently small that the field is constant over it.   For the electric dipole we take the full expressions for its potential, field, and charge density, namely \cite{GSPON2004D}
\begin{eqnarray}
\label{ele:1}
                \phi_{dm}(\vec{r})
         &=   \dfrac{1}{r^2} \vec{p} \cdot \vec{u} \UPS(r),\\
\label{ele:2}
    \vec{E}_{dm}(\vec{r}) 
         &= \Bigl( 3\frac{\vec{u}(\vec{p} \cdot \vec{u})}{r^3}
          - \frac{\vec{p}}{r^3} \Bigr)  \UPS(r)
  -  \frac{\vec{u}(\vec{p}\cdot\vec{u})}{r^2} \DUP(r),\\
\label{ele:3}
         \rho_{dm}(\vec{r}) 
         &= \dfrac{3}{4\pi} \frac{ \vec{p} \cdot \vec{u} }{r^3} \DUP(r),
\end{eqnarray}
and for the vector potential, field, and current density of the uniform magnetic field the standard expressions
\begin{eqnarray}
\label{ele:4}
          \vec{A}  &= \dfrac{1}{2} \vec{ B}_0 \times \vec{r},\\
\label{ele:5}
          \vec{B}  &=  \vec{ B}_0,\\
\label{ele:6}
          \vec{j}  &= 0.
\end{eqnarray}
As the current density is identically zero we can immediately conclude that Furry's formula without surface terms, i.e., equation \eqref{int:2}, would give a zero result, which is why we will have to use the general formula \eqref{gen:5} instead.  We also remark that $\vec{A}$ satisfies the Coulomb gauge so that Faraday's formula is applicable in principle.

  Finally, for the integration volume we take a sphere of radius $R$, and since we work in spherical coordinates we only need a few elementary formulae, i.e.,
\begin{eqnarray}
\label{ele:7}
                    d^3V = d\omega~dr~r^2, \quad &\mbox{and}\quad
          d^2\vec{S} = d\omega~\vec{u} R^2,\\
\label{ele:8}
        \iint d\omega~ \vec{u}(\vec{p}\cdot\vec{u})
           = \dfrac{4\pi}{3}\vec{p}, \quad & \mbox{and}\quad
        \iint d\omega~ \vec{u}\times(\vec{p}\times\vec{u})
           = \dfrac{8\pi}{3}\vec{p}.
\end{eqnarray}

  We start with Poynting's formula, equation \eqref{int:1} or \eqref{gen:1}, and replace $\vec{E}$ and $\vec{B}$ by \eqref{ele:2} and \eqref{ele:5} to obtain
\begin{eqnarray}
\label{ele:9}
    \vec{P}_{\rm Poynting}
      &= \iiint d^3V \dfrac{\vec{E}_{dm} \times \vec{B}_0}{4 \pi}\\
\label{ele:10}
      &= -\dfrac{1}{4 \pi} \int_{0}^{R} dr~  \iint d\omega ~\frac{\UPS(r)}{r} 
           \Bigl( 3\vec{u}(\vec{p} \cdot \vec{u})
                  - \vec{p} \Bigr)  
         \times \vec{B}_0\\
\label{ele:11}
      &~~~ -\dfrac{1}{4 \pi} \int_{0}^{R} dr~  \iint d\omega ~\DUP(r) 
         \vec{u}(\vec{p}\cdot\vec{u}) 
         \times \vec{B}_0.
\end{eqnarray}
Ignoring the $\UPS$ factor it is seen that the radial integral in equation \eqref{ele:10} diverges as $1/r$ at $r=0$, and taking $\UPS$ into account leads to $\UPS(r)/r=0/0$ which is undefined.  On the other hand, using \eqref{ele:8} for integrating over the angles, this equation would be zero if the result from the radial integration was finite.  Thus, as the radial and angular integrations cannot be interchanged, the result is ambiguous.  This ambiguity is clearly related to the $1/r$ divergent factor, which is one instance of the fundamental divergencies which plague classical and quantum electrodynamics.  As is well known, there is no other cure than ignoring these divergences, for example by introducing a ``covariant'' form-factor, and possibly ``renormalising'' the result afterwards. In the present case, a consistent method is to represent the $\UPS$ and $\DUP$ functions by the limits $\varepsilon \rightarrow 0$ of a pair of functions such as, for example,
\begin{eqnarray}
\label{ele:12}
    \Upsilon_{\varepsilon}(r) \DEF  
    \frac{2}{\pi} \arctan (\frac{r}{\varepsilon}),
    \qquad\mbox{and}\qquad
    \delta_{\varepsilon}(r) \DEF  
    \frac {2}{\pi} \frac{\varepsilon}{\varepsilon^2+r^2},
\end{eqnarray}
and to keep $\varepsilon$ very small but \emph{finite} so that $\UPS_{\varepsilon}(r)/r=2/\pi\varepsilon$ at $r=0$. In that case the radial integration is finite while angular integration gives zero.

    Equation  \eqref{ele:10} is thus zero, and it remains to calculate equation \eqref{ele:11}, which using \eqref{ele:8} and \eqref{inc:4} gives the finite contribution
\begin{eqnarray}\label{ele:13}
    \vec{P}_{\rm Poynting}
           = \dfrac{1}{3} \vec{B}_0 \times \vec{p},
\end{eqnarray}
which has its origin in the $\DUP$ singularity of the field strength \eqref{ele:2}.

   As everything is finite, we can apply the general Furry formula \eqref{gen:5}, and calculate the surface term using \eqref{ele:1} for $\phi$ and \eqref{ele:5} for $\vec{B}$.  We get
\begin{eqnarray}
\label{ele:14}
    -\dfrac{1}{4\pi} \iint \phi d^2\vec{S} \times \vec{B} =
    -\dfrac{1}{4\pi} \iint \dfrac{1}{R^2} \vec{p} \cdot \vec{u}
                          ~d\omega R^2\vec{u} \times \vec{B_0},
\end{eqnarray}
that is, after angular integration,
\begin{eqnarray}\label{ele:15}
    \vec{P}_{\rm Furry,~Eq.~\eqref{gen:5}}
      = \dfrac{1}{3} \vec{B}_0 \times \vec{p},
\end{eqnarray}
which agrees with \eqref{ele:13}.

   The first surface term in the general Faraday formula \eqref{gen:16} contains the expression $\underline{\overrightarrow{\nabla} (\vec{A}}\cdot d^2\vec{S})$ which has to be evaluated assuming that the surface element is constant.  Using \eqref{ele:4} for $\vec{A}$ we therefore calculate 
\begin{eqnarray}
\label{ele:16}
  \underline{\overrightarrow{\nabla} (\vec{A}}\cdot \vec{C})
     = (\vec{C} \cdot \nabla)\vec{A} 
                  + \vec{C}\times (\nabla \times  \vec{A})\\
\label{ele:17}
  = (\vec{C} \cdot \nabla)\frac{1}{2}\vec{B}_0 \times \vec{r}
                  + \vec{C}\times \vec{B}_0\\
\label{ele:18}
  = (\vec{C} \cdot \vec{u})
     \frac{1}{2}\vec{B}_0 \times \vec{u}
                  + \vec{C}\times \vec{B}_0.
\end{eqnarray}
Thus, replacing the constant vector $\vec{C}$ by $d\omega R^2\vec{u}$,
\begin{eqnarray}
\label{ele:19}
     \underline{\overrightarrow{\nabla} (\vec{A}}\cdot d^2\vec{S})
     = d\omega R^2 \frac{1}{2} \vec{u} \times \vec{B}_0.
\end{eqnarray}
We now use \eqref{ele:1} for $\phi$ and find
\begin{eqnarray}
\nonumber
            -\dfrac{1}{4\pi} \iint 
 \phi\underline{\overrightarrow{\nabla} (\vec{A}}\cdot d^2\vec{S})
         &= -\dfrac{1}{4\pi} \iint 
       \dfrac{1}{R^2} \vec{p} \cdot \vec{u}
      d\omega R^2 \frac{1}{2} \vec{u} \times \vec{B}_0\\
\label{ele:20}
         &= \dfrac{1}{6} \vec{B}_0 \times \vec{p}.
\end{eqnarray}
The calculation of the second surface term in \eqref{gen:16} is straightforward: using \eqref{ele:2} for $\vec{E}$ and \eqref{ele:4} for $\vec{A}$ we get
\begin{eqnarray}
\nonumber
       -\dfrac{1}{4\pi} \iint (\vec{E}\cdot d^2\vec{S}) \vec{A} 
    &= -\dfrac{1}{4\pi} \iint d\omega  \dfrac{R^2}{R^3}
               \Bigl[ \Bigl( 3\vec{u}(\vec{p}\cdot\vec{u})
       - \vec{p} \Bigr)
       \cdot \vec{u}\Bigr] \frac{1}{2}\vec{B}_0 \times \vec{r}\\
\label{ele:21}
    &= -\dfrac{1}{4\pi} \iint d\omega (\vec{p}\cdot\vec{u})
                                      \vec{B}_0 \times \vec{u}
     = - \dfrac{1}{3} \vec{B}_0 \times \vec{p}.
\end{eqnarray}
We finally need the volume term in \eqref{gen:16}, which is also straightforward to calculate: using \eqref{ele:3} for $\rho$ and \eqref{ele:4} for $\vec{A}$ we obtain
\begin{eqnarray}\label{ele:22}
    \iiint d^3V \rho \vec{A}
      = \int_{0}^{R} dr~\iint d\omega~ r^2 
    \dfrac{3}{4\pi}
    \frac{ \vec{p} \cdot \vec{u}}{ r^3 } \DUP(r)
    \dfrac{1}{2} \vec{B}_0 \times \vec{ r} 
     = \dfrac{1}{2} \vec{B}_0 \times \vec{p}.
\end{eqnarray}
It remains to add \eqref{ele:20}, \eqref{ele:21}, and \eqref{ele:22}, i.e., $(1/6-1/3+1/2)\vec{B}_0 \times \vec{p}$, to get in total
\begin{eqnarray}\label{ele:23}
    \vec{P}_{\rm Faraday,~Eq.~\eqref{gen:16}}
      = \dfrac{1}{3} \vec{B}_0 \times \vec{p},
\end{eqnarray}
in agreement with \eqref{ele:13} and \eqref{ele:15}.

  In conclusion, we have found that by including the surface terms the general Furry and Faraday formulae give the same electromagnetic momentum as calculated with Poyntings formula, provide the $\DUP$-function singularity that is present in the dipolar electric field $\vec{E}_{dm}$ is included, whereas the $1/r$ singularity in the electromagnetic momentum is discarded. 

   As an application of the result obtained in this section, we assume that the volume in which the magnetic field is constant is the gap between the plates of a capacitor, which it-self is filled by a dielectric material such that its total electric polarisation is $\vec{p}$:  The total electromagnetic momentum in the dielectric material is then $\vec{B}_0 \times \vec{p}/3$.

  As another application we transpose our result for an electric dipole in a magnetic field to the case of a magnetic dipole in a locally uniform electric field.  Indeed, if we consider Poynting's formula \eqref{ele:9} and replace $\vec{E}_{dm}$ by $\vec{E}_0$, and $\vec{B}_0$ by $\vec{B}_{d}$, we see by comparing \eqref{int:4} and \eqref{int:6} that the only difference is in the  angular dependence of the $\DUP$-function singularity.  For a magnetic dipole of moment $\vec{m}$ we have therefore
\begin{eqnarray}\label{ele:24}
    \vec{P}_{\rm Poynting}
           = \dfrac{2}{3} \vec{E}_0 \times \vec{m},
\end{eqnarray}
where the factor 2/3 comes from the angular integration with \eqref{ele:8} of the double cross product in \eqref{int:4}.

\section{Magnetic dipole in the field of a point charge}
\label{mag:0} \setcounter{equation}{0}

We now consider a magnetic dipole located at $\vec{r} = 0$, and a point charge positioned at $\vec{r} = \vec{a}$, an example considered by Furry in \cite{FURRY1969-}.  For the integration volume we take the whole space, and for the potentials, fields, and current densities the complete expressions including all singularities \cite{GSPON2004D}.  Thus, for the point charge we have
\begin{eqnarray}       
\label{mag:1}
          \phi_p(\vec{r}) &= q\frac{1}{|\vec{r}-\vec{a}|}
                                      \UPS(|\vec{r}-\vec{a}|),\\
\label{mag:2}
 \vec{E}_p(\vec{r}) &= q\frac{\vec{r}-\vec{a}}
                                    {|\vec{r}-\vec{a}|^3}
                                \UPS(|\vec{r}-\vec{a}|) 
                             - q\frac{\vec{r}-\vec{a}}
                                    {|\vec{r}-\vec{a}|^2}
                                \DUP(|\vec{r}-\vec{a}|),\\
\label{mag:3}
          \rho_p(\vec{r}) &= q\frac{1}{4\pi |\vec{r}-\vec{a}|^2}
                                           \DUP(|\vec{r}-\vec{a}|),
\end{eqnarray}
and for the magnetic dipole
\begin{eqnarray}
\label{mag:4}
                 \vec{A}_{d}(\vec{r})
      &=   \dfrac{\vec{m} \times \vec{u}}{r^2}  \UPS(r),\\
\label{mag:5}
                 \vec{B}_{d}(\vec{r})
      &= \Bigl( 3\frac{\vec{u}(\vec{m} \cdot \vec{u})}{r^3}
          - \frac{\vec{m}}{r^3} \Bigr)  \UPS(r)
          + \frac{\vec{u} \times
                     (\vec{m} \times \vec{u})}{r^2} \DUP(r),\\
\label{mag:6}
                 \vec{j}_{d}(\vec{ r}) 
      &= \dfrac{3}{4\pi} \frac{ \vec{m} \times \vec{u} }
                                { r^3 } \DUP(r),
\end{eqnarray}

  We start with Poynting's formula, equation \eqref{int:1} or \eqref{gen:1}, and replace $\vec{E}$ and $\vec{B}$ by \eqref{mag:2} and \eqref{mag:5}. Assuming $|\vec{a}| \neq 0$ we have $\DUP(|\vec{r}-\vec{a}|)\DUP(r) = 0$ so that the product of the two $\DUP$ terms is zero.  Under the same assumption the product of the $\DUP(|\vec{r}-\vec{a}|)$ and $\UPS(r)$ terms is also zero.  Indeed, after recentering the coordinate system at $\vec{r} =\vec{a}$ we get as radial integral $\int dr~ r\DUP(r)=0$. We are thus left with two terms:
\begin{eqnarray}
\label{mag:7}
     \fl \qquad \qquad
    \vec{P}_{\rm Poynting}
      &= \iiint d^3V \dfrac{\vec{E}_{p} \times \vec{B}_d}{4 \pi}\\
\label{mag:8}
     \fl \qquad \qquad
      &= \dfrac{1}{4 \pi} \int_{0}^{\infty} dr  \iint d\omega~
                               q\frac{\vec{r}-\vec{a}}
                                    {|\vec{r}-\vec{a}|^3}
                                \UPS(|\vec{r}-\vec{a}|) \times
                                                         \frac{\UPS(r)}{r} 
           \Bigl( 3\vec{u}(\vec{p} \cdot \vec{u})
                  - \vec{p} \Bigr)  \\
\label{mag:9}
     \fl \qquad \qquad
      &+ \dfrac{1}{4 \pi} \int_{0}^{\infty} dr  \iint d\omega ~
                               q\frac{\vec{r}-\vec{a}}
                                    {|\vec{r}-\vec{a}|^3}
                                \UPS(|\vec{r}-\vec{a}|) \times \DUP(r) 
              \Bigl( \vec{u} \times (\vec{m} \times \vec{u}) \Bigr).
\end{eqnarray}
Contrary to what happened in the previous section's equation \eqref{ele:10}, the angular integration in \eqref{mag:8} does not give zero because $|\vec{r}-\vec{a}|\neq|\vec{r}+\vec{a}|$.  Moreover, there are two singular points to consider: $\vec{r}=\vec{a}$ and $\vec{r}=0$.  Fortunately, the point $\vec{r}=\vec{a}$ gives no trouble because the volume element in spherical coordinates centered at this point has a factor $|\vec{r}-\vec{a}|^2$ which cancels the factor $|\vec{r}-\vec{a}|^{-2}$ in $\vec{E}_p$.  

As for the $1/r$ divergence, it can be taken care of in the same way as was done in the previous section with equation \eqref{ele:10}.  The reason is that very close to $\vec{r}=0$ the difference between $|\vec{r}-\vec{a}|$ and $|\vec{r}+\vec{a}|$ becomes negligible and the net result of angular integration in a region close to $\vec{r}=0$ is to yield a zero provided the radial integral over that region is finite.  Thus, if we replace the zero lower bound in \eqref{mag:8} by a small value $\varepsilon$ we get a finite result, which moreover is independent of the precise value of $\varepsilon$.  Unfortunately the evaluation of this integral is tedious, so that we just quote the result of reference \cite[p.~624--625]{FURRY1969-}, i.e.,
\begin{eqnarray}\label{mag:10}
    \vec{P}_{\rm Poynting, Eq.~\eqref{mag:8}}
      = \dfrac{1}{3} \dfrac{q}{a^3} \vec{m} \times \vec{a}.
\end{eqnarray}
On the other hand, the singular contribution \eqref{mag:9} is easy to calculate because $\DUP(r)$ implies that we can replace  $\vec{r}-\vec{a}$ by $-\vec{a}$.  Thus,
\begin{eqnarray}
\nonumber
    \vec{P}_{\rm Poynting, Eq.~\eqref{mag:9}}
      &= -\dfrac{1}{4 \pi} \int_{0}^{\infty} dr~\DUP(r) \iint d\omega ~
                     q\frac{\vec{a}}{a^3} \times 
        \Bigl( \vec{u} \times (\vec{m} \times \vec{u}) \Bigr)\\
\label{mag:11}
      &= \dfrac{2}{3} \dfrac{q}{a^3} \vec{m} \times \vec{a}.
\end{eqnarray}
This partial result has a simple interpretation: It is the electromagnetic momentum of a magnetic dipole at a point where the electric field is  $\vec{E}_0 = -q\vec{a}/a^3$, i.e., the electric field at the origin in our configuration. Thus equation \eqref{mag:11} agrees with \eqref{ele:24}. 

  Adding \eqref{mag:10} and \eqref{mag:11} we obtain the total electromagnetic momentum of a system comprising a point magnetic dipole and a point charge, i.e.,
\begin{eqnarray}\label{mag:12}
    \vec{P}_{\rm Poynting}
      =  \dfrac{q}{a^3} \vec{m} \times \vec{a}
\end{eqnarray}
This result is easily confirmed by means of Furry's and Faraday's formulae in the simple forms \eqref{int:1} and \eqref{int:2} because the integrands in the surface terms fall of more rapidly than  $r^{-3}$ for $r \rightarrow \infty$, and because the vector potential \eqref{mag:4} satisfies the Coulomb gauge.

For instance, Furry's formula is
\begin{eqnarray}\label{mag:13}
    \vec{P}_{\rm Furry} =  \iiint d^3V \phi \vec{j}  =  \iiint d^3V
                       q\frac{1}{|\vec{r}-\vec{a}|}
                            \UPS(|\vec{r}-\vec{a}|)
    \dfrac{3}{4\pi} \frac{ \vec{m} \times \vec{u} }{r^3} \DUP(r),
\end{eqnarray}
which after developing $1/|\vec{r}-\vec{a}|$ around $\vec{r} = 0$ as
\begin{eqnarray}\label{mag:14}
    \frac{1}{|\vec{r}-\vec{a}|} \cong
    \frac{1}{a}\bigl(1 + \frac{\vec{r}\cdot\vec{a}}{a^2}\bigr),
\end{eqnarray}
gives a result equal to \eqref{mag:12}.  Similarly, Faraday's formula is
\begin{eqnarray}\label{mag:15}
    \vec{P}_{\rm Faraday} =  \iiint d^3V \rho \vec{A} =  \iiint d^3V
      q\frac{\DUP(|\vec{r}-\vec{a}|)}{4\pi |\vec{r}-\vec{a}|^2}
                    \dfrac{\vec{m} \times \vec{u}}{r^2} \UPS(r),
\end{eqnarray}
which after the change of integration variable $r \rightarrow |\vec{r}-\vec{a}|$  also gives a total electromagnetic momentum equal to \eqref{mag:12}.  We have therefore confirmed the total \eqref{mag:12} by means of Furry's and Faraday's formulae, which by subtracting the partial result \eqref{mag:11} enables us to confirm the correctness of Furry's independent calculation of the first partial result \eqref{mag:10}.  

It is now worthwhile to compare the method used here to that of Furry \cite{FURRY1969-}: In our case we have assumed that the dipole was strictly point-like so that we had to use a regularisation process typical of quantum electrodynamics to discard the unphysical $1/r$ singularity in \eqref{mag:8}, while the physical $\DUP$ singularity in \eqref{mag:5} led to the finite $\DUP$-function contribution given by \eqref{mag:9}.  On the other hand, Furry assumed that close to $r=0$ the point-like dipolar field had to be replaced by that of a magnetised sphere of very small radius, which gave a contribution equal to that of our $\DUP$ singularity. 

Finally, it is of interest to rewrite equation \eqref{mag:12} in the form
\begin{eqnarray}\label{mag:16}
    \vec{P}_{\rm Poynting}
      =  \vec{E}_0 \times \vec{m},
\end{eqnarray}
where $\vec{E}_0 = -q\vec{a}/a^3$ is the electric field of the point charge at the position of the magnetic dipole.  Indeed, this expression immediately generalises to the case where any static charge distribution is the sole source of $\vec{E}_0$, a result obtained by Aharonov \emph{et~al}., i.e., \cite[Eq.~(4)]{AHARO1988-}, using a variant of the Furry formula \cite[Eq.~(46)]{FURRY1969-}.  Thus, \eqref{mag:16} is the total electromagnetic momentum, i.e., the so-called ``hidden momentum,'' of any configuration comprising a point magnetic dipole in a constant electric field.

\section{Finite electric and magnetic dipolar system}
\label{fin:0} \setcounter{equation}{0}

In this section we consider a finite electric and magnetic dipolar system made of two concentric shells, one of radius $a$ that is ``dipole charged,'' and another of radius $b > a$ that is ``dipole magnetised'' \cite{ROMER1995-}.  The outer shell has a surface current distribution producing a magnetic field that is dipolar outside it and uniform inside, while the inner shell has a surface charge distribution producing an electric field that is dipolar outside and uniform inside.  While this configuration is unrealistic in the sense that it would be very difficult to actually build it \cite{HNIZD1997-}, it is perfectly consistent with the laws of electrodynamics, and has the advantage that all calculations are relatively simple. Moreover, for the purpose of calculating the electromagnetic momentum, this system is similar to a spherical capacitor immersed in a uniform magnetic field such as considered in \cite{MCDON2006-}.

\subsection{Dipolar fields of charge/current carrying shells}

The fields generated by dipolar distributions on shells provide an example of how the distributional methods of section \ref{inc:0} can be used in a case where the singularities are spherical rather than punctual, because by symmetry they corresponds to a single point of the radial variable $r$.

   For instance, the scalar potential, field strength, and charge density distribution of a charged-shell of radius $a$ producing a constant electric field inside it and a dipolar electric field outside are
\begin{eqnarray}
\label{fin:1}
                 \phi_{dm}(\vec{r},r \geqslant a)
      &=  \dfrac{\vec{p} \cdot \vec{u}}{r^2}  \UPS(r-a),\\
\label{fin:2}
                 \vec{E}_{dm}(\vec{r},r \geqslant a)
      &= \Bigl( 3\frac{\vec{u}(\vec{p} \cdot \vec{u})}{r^3}
          - \frac{\vec{p}}{r^3} \Bigr)  \UPS(r-a)
  -  \frac{\vec{u}(\vec{p}\cdot\vec{u})}{r^2} \DUP(r-a),\\
\label{fin:3}
                 \rho_{dm}(\vec{ r},r \geqslant a) 
      &= \dfrac{3}{4\pi} \frac{ \vec{p} \cdot \vec{u}}
                                { r^3 } \DUP(r-a).
\end{eqnarray}
As can be seen, these fields go over to equations (\ref{ele:1}--\ref{ele:3}) as $a \rightarrow 0$.

  Similarly, the vector potential, field, and current density distribution on a conducting-shell of radius $b$ producing a constant magnetic field inside it and a dipolar magnetic field outside are
\begin{eqnarray}
\label{fin:4}
                 \vec{A}_{d}(\vec{r},r \geqslant b)
      &=   \dfrac{\vec{m} \times \vec{u}}{r^2}  \UPS(r-b),\\
\label{fin:5}
                 \vec{B}_{d}(\vec{r},r \geqslant b)
      &= \Bigl( 3\frac{\vec{u}(\vec{m} \cdot \vec{u})}{r^3}
          - \frac{\vec{m}}{r^3} \Bigr)  \UPS(r-b)
          + \frac{\vec{u} \times
                     (\vec{m} \times \vec{u})}{r^2} \DUP(r-b),\\
\label{fin:6}
                 \vec{j}_{d}(\vec{ r},r \geqslant b) 
      &= \dfrac{3}{4\pi} \frac{ \vec{m} \times \vec{u} }
                                { r^3 } \DUP(r-b),
\end{eqnarray}
which reduces to the fields of a point magnetic dipole when $b \rightarrow 0$, i.e., (\ref{mag:4}--\ref{mag:6}).  

   Equations (\ref{fin:1}--\ref{fin:6}) are the expressions for the fields in the regions $r \geqslant a$ and r $\geqslant b$ where they have a \emph{dipolar} character.  As is well-known, the corresponding surface charge and current distributions, equations \eqref{fin:3} and \eqref{fin:6}, produce fields that are \emph{uniform} in the regions $r \leqslant a$ and r $\leqslant b$, i.e., inside the shells.\footnote[1]{The proof of this basic result of potential theory requires however a development in Legendre Polynomials or spherical harmonics, as is done, e.g., for of a magnetised sphere in \cite[p.~195]{JACKS1975-}.}  To obtain the expressions of these fields including their singularities we can proceed the same way as we have done for  (\ref{fin:1}--\ref{fin:6}), that is to related these singularities to those of a point source by a substitution of the type $r \rightarrow R-r$ where $R$ is the shell radius, because in spherical symmetry such a substitution is equivalent to just translating a point along the $r$ axis.

Thus, to find the corresponding expression for the uniform electric field generated by a dipolar surface charge distribution we start from the well-known scalar potential of a uniform electric field, namely $\phi(\vec{r}) = - \vec{E}_0 \cdot \vec{u}\, r\UPS(r)$  where we have included the $\UPS$ function taking care of the singularity at $|\vec{r}|=0$.  Substituting $a-r$ for $r$ in the argument of $\UPS$ gives the potential of the electric field, minus the gradient of this potential yields the electric field strength, etc. Thus, remembering that $\UPS(a-r)' = -\DUP(a-r)$,
\begin{eqnarray}
\label{fin:7}
                 \phi_{dm}(\vec{r},r \leqslant a)
      &= - \vec{E}_0 \cdot \vec{u}\, r\UPS(a-r),\\
\label{fin:8}
                 \vec{E}_{dm}(\vec{r},r \leqslant a)
      &= \vec{ E}_0 \UPS(a-r) - \vec{u}
              (\vec{ E}_0 \cdot \vec{u})~ r \DUP(a-r),\\
\label{fin:9}
                 \rho_{dm}(\vec{ r},r \leqslant a) 
      &=  -\dfrac{3}{4\pi} \vec{E}_0 \cdot \vec{u} \DUP(a-r).
\end{eqnarray}

  Similarly, the vector potential, field strength, and current density distribution of the constant magnetic field inside a shell of radius $b$ producing a dipolar magnetic field outside it are obtained from the potential \eqref{ele:4} rewritten as $\vec{A}(\vec{r}) = \vec{B}_0 \times \vec{u} r\UPS(r)/2$ by substituting $b-r$ for the argument of the $\UPS$ function, i.e.,
\begin{eqnarray}
\label{fin:10}
                 \vec{A}_{d}(\vec{r},r \leqslant b)
      &=        \dfrac{1}{2} \vec{B}_0 \times \vec{u} r\UPS(b-r),\\
\label{fin:11}
                 \vec{B}_{d}(\vec{r},r \leqslant b)
      &=   \vec{ B}_0 \UPS(b-r)
       -   \dfrac{1}{2} \vec{u} \times 
          (\vec{B}_0 \times \vec{u})~ r \DUP(b-r),\\
\label{fin:12}
                 \vec{j}_{d}(\vec{ r},r \leqslant b) 
      &=    +\dfrac{3}{8\pi} \vec{ B}_0 \times \vec{u} \DUP(b-r).
\end{eqnarray}

    It remains to relate $\vec{E}_0$ to $\vec{p}$, as well as $\vec{ B}_0$ to $\vec{m}$.  This is done by requiring that the potentials $\phi_{dm}(\vec{r},r > a)$ and $\phi_{dm}(\vec{r},r < a)$, as well as $\vec{A}_{d}(\vec{r},r > b)$ and $\vec{A}_{d}(\vec{r},r < b)$, are continuous at $R=a,b$, respectively.  This gives
\begin{eqnarray}
\label{fin:13}
     \vec{p} = - a^3 \vec{ E}_0,\\
\label{fin:14}
     \vec{m} = \dfrac{1}{2} b^3 \vec{ B}_0.
\end{eqnarray}

We now interpret the fields (\ref{fin:1}--\ref{fin:12}) as those of a finite spherical electric dipole of radius $a$ immersed in a uniform magnetic field produced by an ideal spherical dipolar magnet of radius $b > a$.  However, before applying Poyntings, Furry's, or Faraday's formulae, we need to verify that their conditions of applicability are satisfied.   If we exclude the points on the respective shells at $r=a$ and $r=b$, this is clearly the case for all fields (\ref{fin:1}--\ref{fin:12}) since they are finite and of class $\mathcal{C}^1$ at all other points.  This means that we can divide the whole space into three concentric domains and calculate separately the electromagnetic momenta by all three methods in each of them.  If it turns out that the contributions of the surface terms in the general Furry and Faraday formulae cancel each other at the boundaries $r=a$ and $r=b$, the contributions from all three formulae will add up to the same total, provided the volume integrals do not give problems.  To verify that explicitly it would be necessary to calculate all these surface terms, what we will not do because from the form of the fields we expect them to cancel, just like we do not expect any contribution from the surface terms at infinity since all integrands fall off more rapidly than $r^{-3}$ for $r \rightarrow \infty$.  Referring to the derivation of Furry's and Faraday's formulae made in reference \cite{JACKS2006-}, which implicitly assumes that the fields vanish sufficiently fast at infinity, we also remark that the boundary terms at finite $r$ should not give any problem.  It remains therefore to calculate the volume terms.

\subsection{Furry's expression}

With start with Furry's method because it is the easiest to use since $b > a$ in our configuration.  Thus, using \eqref{fin:1} for $\phi$ and \eqref{fin:6} for $\vec{j}$, we get\footnote[1]{One could equally well have used \eqref{fin:12} for  $\vec{j}$.}
\begin{eqnarray}\label{fin:15}
     \vec{P}_{\rm Furry} = \int_{0}^{\infty} dr~\iint d\omega~ r^2 ~
        \dfrac{\vec{p} \cdot \vec{u}}{r^2}  \UPS(r-a)
        \dfrac{3}{4\pi} \frac{ \vec{m} \times \vec{u} }
                                { r^3 } \DUP(r-b),
 \end{eqnarray}
which becomes, after performing the angular integration using \eqref{ele:8} and expressing $\vec{m}$ in terms of $\vec{ B}_0$ with \eqref{fin:14},
\begin{eqnarray}\label{fin:16}
      \vec{P}_{\rm Furry} = 
          \frac{1}{2}\vec{B}_0 \times \vec{ p}
          \int_{b}^{\infty} dr~
          \dfrac{b^3}{r^3} \UPS(r-a) \DUP(r-b),
 \end{eqnarray}
where we replaced the lower integration bound by $b$ because $\vec{ j}$ is zero for $r<b$.  Since $\UPS(r-a)=1$ for $b > a$, using \eqref{inc:4} we get, therefore,
\begin{eqnarray}\label{fin:17}
      \vec{P}_{\rm Furry} = 
          \frac{1}{2}\vec{B}_0 \times \vec{ p},
\end{eqnarray}
which is the correct value expected for a macroscopic dipole or capacitor \cite{MCDON2006-, HNIZD1997-,ROMER1995-}.

\subsection{Faraday's expression}

Here we use \eqref{fin:3} for $\rho$ and, since $r \leqslant b$, equation \eqref{fin:10} for $\vec{A}$, that is the vector potential inside the magnetic shell where $\vec{B}$ is a constant.  Therefore\footnote[1]{One could equally well have used \eqref{fin:9} for  $\rho$.}
\begin{eqnarray}\label{fin:18}
    \vec{P}_{\rm Faraday} = \int_{0}^{\infty} dr~\iint d\omega~ r^2 
    \dfrac{3}{4\pi}
    \frac{ \vec{p} \cdot \vec{u}}{ r^3 } \DUP(r-a)
    \dfrac{1}{2} \vec{B}_0 \times \vec{ r} \, \UPS(b-r),
\end{eqnarray}
which, after angular integration, becomes
\begin{eqnarray}\label{fin:19}
      \vec{P}_{\rm Faraday} = 
          \frac{1}{2}\vec{B}_0 \times \vec{ p}
          \int_{a}^{b} dr~ \DUP(r-a) \UPS(b-r),
 \end{eqnarray}
where we replaced the lower integration bound by $a$ because $\rho$ is zero for $r<a$, and the upper bound by $b$ since we use \eqref{fin:10} for $\vec{A}$.  As $\UPS(b-r)=1$ for $b > a$, using \eqref{inc:4} we get, therefore,
\begin{eqnarray}\label{fin:20}
      \vec{P}_{\rm Faraday} = 
          \frac{1}{2}\vec{B}_0 \times \vec{ p},
\end{eqnarray}
which is again the correct value expected for a macroscopic dipole or capacitor.

\subsection{Poynting's expression}

To evaluate Poynting's expression one has to integrate the cross products of the four fields \eqref{fin:2}, \eqref{fin:5}, \eqref{fin:8}, and \eqref{fin:11}, which are not very simple due to the $\UPS$ and $\DUP$ singularities at $a$ and $b$.  Moreover, Poynting's expression has to be evaluated in the three regions defined by the radii $0, a, b$, and $\infty$.  Thus, compared to Faraday's and Furry's methods, the full calculation with Poynting's method is laborious. For this reason we relegate the details of these calculations to the appendix, section \ref{app:0}, and just collect their results in this section.

   However, as they are very simple to calculate, we first give the respective contributions to the electromagnetic momentum in the three regions for the case where the singularities are completely \emph{ignored}, that is, when the integrations are made \emph{excluding} the $\UPS$ and $\DUP$ singularities at the radii $a$ and $b$:
\begin{eqnarray}
\label{fin:21}
     \vec{P}_{\rm Poynting}(< a) = 
   \iiint d^3V \dfrac{\vec{E}_0 \times \vec{B}_0}{4 \pi}
                        &=\frac{1}{3} \vec{B}_0 \times \vec{p},\\
\label{fin:22}
     \vec{P}_{\rm Poynting}(a < r < b) =
   \iiint d^3V \dfrac{\vec{E}_{dm} \times \vec{B}_0}{4 \pi}
                        &= 0,\\
 \label{fin:23}
     \vec{P}_{\rm Poynting}( > b) =
   \iiint d^3V \dfrac{\vec{E}_{dm} \times \vec{B}_d}{4 \pi}
                        &=\frac{1}{6}\vec{B}_0 \times \vec{p}.
\end{eqnarray}
Thus, the non-zero contributions add up to $\frac{1}{2}\vec{B}_0 \times \vec{p}$ as they should, and the contribution from the fields located between the polarised and magnetised shells is zero.  Also, referring to the calculation made in section {\ref{ele:0} for a point-dipole, we see that the subtotal $\vec{B}_0 \times \vec{p}/3$ for the region $0 < r < b$ corresponds to what was obtained in that section after taking care of a divergence at $r=0$.  This divergence does not arise here because the point-dipole is replaced by a dipole-polarised shell of finite radius, which implies that the contribution in the region $0 < r < a$ is finite, and that the contribution for $a < r < b$ is zero after angular integration.

   Now, recalling what was said in section \ref{inc:0} about singularities,  we remark that ignoring the $\UPS$ and $\DUP$ singularities at the radii $a$ and $b$ corresponds to the ``electrical engineering perspective'' in which they are considered as spurious.  On the other hand, anticipating on the interpretation that will be put forward in the discussion section, we can also say that discarding these singularities corresponds to the ``field perspective'' in which the electromagnetic momentum is calculated by integrating the Poynting vector over the whole space \emph{excluding} the points at which it is singular, that is the points at which the sources reside. 

   Therefore, to find the physical picture corresponding to the opposite perspective, i.e., the ``particle perspective'' in which the singularities in the fields are \emph{included}, one has to calculate the momenta in such a way that the respective contributions from the regular and singular parts of the fields can be isolated from each other, and this separately in the three regions $0 \leqslant  r \leqslant a$, $a \leqslant  r \leqslant b$, and $b \leqslant  r \leqslant \infty$.

   Consequently, to see what happens when the boundaries $a$ and $b$ are included as upper or lower integration bounds, we write them as $a \pm \alpha$ and $b \pm \beta$, where $\alpha \ll a$ and $\beta \ll b$ are positive infinitesimals, so that the radial part of the integrations made in the appendix are defined as follows:
\begin{eqnarray}\label{fin:24}
    \int_0^{a-\alpha}         dr~(...),  \quad\mbox{}\quad 
    \int_{a+\alpha}^{b-\beta} dr~(...),  \qquad\mbox{and}\qquad 
    \int_{b+\beta}^\infty     dr~(...)  . 
\end{eqnarray}
Thus, as long as the parameters $\alpha$ and $\beta$ are non-zero these integrals give the contributions from the fields ignoring the singularities, and when  $\alpha$ or $\beta$ are set to zero they give these contributions including the singularities.  More precisely, taking the limits $\alpha \rightarrow 0$ and $\beta \rightarrow 0$ corresponds to ignoring the singularities, i.e., the discontinuities at $a$ and $b$, whereas setting $\alpha=0$, and/or $\beta=0$, corresponds to including the contributions from these discontinuities.

   We now look at the results of these integrations, equations \eqref{app:11}, \eqref{app:22}, and \eqref{app:31} of the appendix,  i.e., 
\begin{eqnarray}
\label{fin:25}
     \vec{P}_{\rm Poynting}(0 \leqslant  r \leqslant a-\alpha) 
                             &= \vec{B}_0 \times \vec{p}~
 \frac{1}{3} \frac{r^3}{a^3} \UPS(a-r) \Bigr|_0^{a - \alpha},\\
\notag
     \vec{P}_{\rm Poynting}(a+\alpha \leqslant r \leqslant b-\beta)
                             &= \vec{B}_0 \times \vec{p}~
                 \frac{1}{3}~~~~ \UPS(r-a) \Bigr|_{a+\alpha}^{b-\beta}\\
\label{fin:26}
                             &- \,\vec{B}_0 \times \vec{p}~
                 \frac{1}{6}~~~~ \UPS(b-r) \Bigr|_{a+\alpha}^{b-\beta},\\
  \label{fin:27}
     \vec{P}_{\rm Poynting}(b+\beta \leqslant r \leqslant \infty)
                             &=  \vec{B}_0 \times \vec{p}~
 \frac{1}{6} \frac{b^3}{r^3} \UPS(r-b)  \Bigr|_{b+\beta}^\infty.
\end{eqnarray}
For instance, when both $\alpha \neq 0$ and $\beta \neq 0$ the $\UPS$ functions in (\ref{fin:25}--\ref{fin:27}) are all equal to $\UPS(x\neq 0)=1$, which implies in particular that \eqref{fin:26} is zero. Letting $\alpha \rightarrow 0$ and $\beta \rightarrow 0$ we recover the results (\ref{fin:21}--\ref{fin:23}), that is the electromagnetic momenta when the singularities are ignored.  In this case we can also say that the distributional content of (\ref{fin:25}--\ref{fin:27}), which quantifies the discontinuous effects of the singularities at $a$ and $b$, is ignored.

On the other hand, if either $\alpha$ or $\beta$ is set to zero, equations (\ref{fin:25}--\ref{fin:27}) must be interpreted in their distributional sense, which means that the $\UPS$ functions at $a$ or $b$ take their point-values of $\UPS(0)=0$ when $\alpha$ or $\beta$ is zero, as stipulated  by \eqref{inc:2}.   In particular, whereas \eqref{fin:25} and \eqref{fin:27} show an $r$-dependence that is continuous through the factors $r^{\pm 3}$, equation \eqref{fin:26} is strictly discontinuous and thus gives a non-zero contribution only when the integration bounds are exactly equal to $a$ and/or $b$.

For example, when $\alpha=0$ the electromagnetic momentum given by \eqref{fin:25} for $0 \leqslant  r \leqslant a$ is zero: the contribution from the  singularity at $a$ cancels the contribution of the integral in the interval $0 \leqslant  r < a$.  At the same time the integral \eqref{fin:26} is no more zero but yields a finite contribution of $\vec{B}_0 \times \vec{p}/3$ at the bound $r=a$, that is, exactly the contribution \eqref{fin:25} gave when the singularity at $a$ was ignored:  the electromagnetic momentum which was considered as being in the field between 0 and $a$ has been ``concentrated'' at the singularity at $a$.

The same analysis can be repeated for the singularity at $r=b$, with the conclusion that the effect of including or excluding the contribution of that singularity is to concentrate an electromagnetic momentum of $\vec{B}_0 \times \vec{p}/6$ at $r=b$, or to leave it at $r>b$.  

  In conclusion, the total electromagnetic momentum is always $\vec{B}_0 \times \vec{p}/2$, and the effect of including/excluding the singularities is to transfer part or all of that momentum to their locations, or to leave it in the fields.  If we now simplify (\ref{fin:25}--\ref{fin:27}) by taking the foregoing analysis into account, we are led to the following equations:
\begin{eqnarray}
\label{fin:28}
     \vec{P}_{\rm Poynting}(0 \leqslant  r \leqslant a-\alpha) 
                          &= \vec{B}_0 \times \vec{p}~
          \frac{1}{3}\UPS(\alpha),\\
\label{fin:29}
     \vec{P}_{\rm Poynting}(a+\alpha)
                          &= \vec{B}_0 \times \vec{p}~
        \frac{1}{3} \Bigl(1-\UPS(\alpha)\Bigr)\\
\label{fin:30}
     \vec{P}_{\rm Poynting}(b-\beta)
                          &= \vec{B}_0 \times \vec{p}~
           \frac{1}{6} \Bigl(1-\UPS(\beta)\Bigr),\\
\label{fin:31}
     \vec{P}_{\rm Poynting}(b+\beta \leqslant r \leqslant \infty)
                         &=  \vec{B}_0 \times \vec{p}~
                            \frac{1}{6}\UPS(\beta).
\end{eqnarray}
Here $\alpha$ and $\beta$ are again small numbers which are either zero or non-zero, but since $\UPS(r)$ is discontinuous at zero their purpose is just to give results equivalent to those of integrals (\ref{fin:25}--\ref{fin:27}) when the singularities are included or not.  In particular, the total of these equations is always $\vec{B}_0 \times \vec{p}/2$, and the splitting of \eqref{fin:26} into \eqref{fin:29} and \eqref{fin:30} stresses that these contributions arise exclusively from the singularities at $a$ or $b$.

\section{Discussion}
\label{con:0} \setcounter{equation}{0}

   The primary goal of this paper was to better understand under which conditions the methods of  Poynting, Furry, and Faraday can be used to calculated the electromagnetic momentum of static charge and steady current distributions, either taken as whole systems, or as composite systems of finite or infinite size.  In order to do that Furry's and Faraday's formulae were generalised to the case of a finite volume, which led to generalised formulae including surface terms, equations \eqref{gen:5} and \eqref{gen:16}, which for that reason do not have the simplicity and elegance of Furry's and Faraday's original formulae \eqref{int:2} and \eqref{int:3}.  Nevertheless, the generalised formulae can, in principle, easily be applied to any system, or part of system, in which the potentials, fields, and sources are continuous and continuously differentiable functions --- that is, to realistic systems which are typically considered by ``electrical engineers.''

   The second main goal of this paper was to calculate the electromagnetic momentum of systems in which the potentials, fields, and sources may have singularities and to use for that purpose the method presented in reference \cite{GSPON2004D}.   As explained in section \ref{inc:0}, that method consists of making explicit in the potentials the discontinuity which through two successive differentiations leads to the singularity corresponding to the sources.

   In section \ref{ele:0}, the electromagnetic momentum of a point dipole in a locally uniform field was calculated with the generalised Furry and Faraday formulae and found to agree with Poynting's formula.  As this calculation required discarding a diverging term, the same formalism was applied in section \ref{mag:0} to a point dipole in the field of a point charge, and shown to be consistent and in agreement with the methods of references \cite{FURRY1969-} and \cite{AHARO1988-}.  In both applications the $\DUP$-function singularity in the field strength of the point-dipoles was found to contribute the \emph{finite} part of the dipoles's contributions to the total momentum after discarding the $1/r$ divergence.  

    In the third application, section \ref{fin:0}, the location of electromagnetic momentum in a composite system made of two concentric shells modelling a spherical electric dipole of radius $a$ immersed in the uniform magnetic field produced by a spherical magnet of radius $b$ was investigated. 

   In sections \ref{fin:0}.2 and \ref{fin:0}.3, the total electromagnetic momentum was easily found to be equal to $\vec{B}_0 \times \vec{p}/2$ using either Furry's or Faraday's methods, which could be applied in a straightforward manner because the finite size of the model insured that  there is no singularity at the origin (contrary to the point-dipole systems considered in sections \ref{ele:0} and \ref{mag:0}). This total momentum was confirmed in section \ref{fin:0}.4 using Poynting's method, which also gave additional information on the location of the electromagnetic momentum that is not easily provided by Furry's and Faraday's methods --- unless the complicated surface terms which cancel each others at the boundaries are explicitly calculated.

   In section \ref{fin:0}.4, the electromagnetic momenta were calculated separately for the three regions $(0 \leqslant  r \leqslant a)$, $(a \leqslant  r \leqslant b)$, and $(b \leqslant r \leqslant \infty)$, and their significance analysed.  The results are presented in detail in equations (\ref{fin:25}--\ref{fin:27}), and in simplified form in equations (\ref{fin:28}--\ref{fin:31}).  In both sets two infinitesimal parameters $\alpha$ and $\beta$ enable us to include or exclude the contributions from the singularities in the field strengths at the radii $a$ and $b$: when $\alpha=0$ or $\beta=0$ the singularities at $r=a$ or $r=b$ are included, whereas when $\alpha\neq 0$ or $\beta\neq 0$ they are excluded.  This further enables us to summarise the results as four  symbolic equations,  (\ref{con:1}--\ref{con:4}), which correspond to the four possible ways the partial results (\ref{fin:28}--\ref{fin:31}) can be added up by either including or not the contributions from the singularities at $a$ and $b$:
\begin{itemize}

\item {\bf Field perspective ($\alpha \neq 0, \beta \neq 0)$ :}
\begin{eqnarray}\label{con:1}
    \int_0^{a-\alpha}         + 
    \int_{a+\alpha}           + 
    \int           ^{b-\beta} + 
    \int_{b+\beta}^\infty     =
    \frac{1}{3} +   0   +   0  + \frac{1}{6} = \frac{1}{2}.
\end{eqnarray}
{\it The electromagnetic momentum is in the fields at $r<a$ and $r>b$.}

\item {\bf Particle perspective ($\alpha = 0, \beta = 0)$ :}
\begin{eqnarray}\label{con:2}
    \int_0^{a}          + 
    \int_{a}            + 
    \int    ^{b}        + 
    \int_{b}^\infty     =
    0 +   \frac{1}{3}   +   \frac{1}{6}   + 0 = \frac{1}{2}.
\end{eqnarray}
{\it The electromagnetic momentum is concentrated at the location of the charge and current distributions at $a$ and $b$.}

\item {\bf ``Faraday's'' perspective ($\alpha = 0, \beta \neq 0)$ :}
\begin{eqnarray}\label{con:3}
    \int_0^{a}             + 
    \int_{a}               + 
    \int    ^{b-\beta}     + 
    \int_{b+\beta}^\infty  =
    0 +  \frac{1}{3}  +  0  + \frac{1}{6} = \frac{1}{2}.
\end{eqnarray}
{\it The electromagnetic momentum is at the location of the charge distribution at $a$, and in the fields at $r>b$.}

\item {\bf ``Furry's'' perspective ($\alpha\neq 0, \beta = 0)$ :}
\begin{eqnarray}\label{con:4}
    \int_0^{a-\alpha}      + 
    \int_{a+\alpha}        + 
    \int           ^{b}    + 
    \int_{b}^\infty        =
    \frac{1}{3} +  0  +  \frac{1}{6}  + 0 = \frac{1}{2}.
\end{eqnarray}
{\it The electromagnetic momentum is at the location of the current distribution at $b$, and in the fields at $r<a$.}

\end{itemize} 
In these symbolic equations the first and fourth integral signs correspond to the integrals in the regions $(0 \leqslant  r \leqslant a)$ and $(b \leqslant r \leqslant \infty)$, respectively; and the two integral signs in between them correspond to the contributions of the integral over the region  $(a \leqslant  r \leqslant b)$, which are non-zero only when the singularities at the lower and/or upper bounds are included.

Mathematically, these four possibilities have their origin in the properties of the singularities of the fields at $a$ and $b$.  The contributions arising from these singularities are such that the integral $P=\iint d\omega\int_{0}^{\infty} dr\,r^2 \vec{E}\times\vec{B}/4\pi$ has a unique value which is the \emph{same} whether any singularity is included or not (just like in complex analysis contributions of poles can be included or excluded by an appropriate choice of the integration path).

In physical language, this corresponds to the well-known ``wave-particle'' duality of field-theory, which means that there is as much information in the ``fields'' than in the ``particles'' \cite{BARUT1993-}, a property that in mathematics is systematically exploited in the theory of complex and hyper-complex functions.  Thus, our result is a beautiful confirmation of Lanczos's intuition, namely that electrodynamics can be interpreted as a four-dimensional generalisation of Cauchy's theory such that singularities correspond to electrons \cite{HURNI2004-}, and is therefore providing another example of the remarkable diversity of mathematical and physical pictures which can be invoked to interpret electromagnetic phenomena.

Finally, the reason why the symbolic equations \eqref{con:3} and \eqref{con:4} have been labeled as ``Faraday's'' and ``Furry's'' is that they suggest an interpretation of the Faraday and Furry formulae with regards to the location of the electromagnetic momentum.  Indeed, in order to integrate over the charge and current distributions as in the these formulae, it is necessary that the singularities in the field strengths have been included (either explicitly as with the method of reference \cite{GSPON2004D}, or implicitly as is usually the case), because otherwise the charge and current distributions would be zero.

\section*{Dedication}

This paper is dedicated to the memory of Val Telegdi (1922--2006), ``doktorvater'' of Kirk McDonald and Andre Gsponer at University of Chicago, who inspired an unequaled spirit of rigour in his students, and who would most certainly have enjoyed this paper.

\section*{Acknowledgements}

The author wishes to thank Vladimir Hnizdo, Jean-Pierre Hurni, and Kirk McDonald for extensive correspondence related to this paper.

\section{Appendix: Poynting's method for finite dipolar system}
\label{app:0} \setcounter{equation}{0}

In this appendix we calculate the electromagnetic momentum with Poynting's method for the double-shell system of section \ref{fin:0}, leaving the radial integrals in parametric form with the upper or lower integration bounds written as $a \pm \alpha$ and $b \pm \beta$, where $\alpha \ll a$ and $\beta \ll b$ are positive infinitesimals.  Thus we calculate the momenta separately in the three regions $0 \leqslant  r \leqslant a$, $a \leqslant  r \leqslant b$, and $b \leqslant  r \leqslant \infty$ according to equation \eqref{int:1}, but leave the radial integrals as given by \eqref{fin:24}, i.e., 
\begin{eqnarray}\label{app:1}
    \int_0^{a-\alpha}         dr~(...),  \qquad\mbox{}\qquad 
    \int_{a+\alpha}^{b-\beta} dr~(...),  \qquad\mbox{and}\qquad 
    \int_{b+\beta}^\infty     dr~(...), 
\end{eqnarray}
This  enables us to find out where the electromagnetic momentum is located in the system, as well as the respective amounts of the various regular and singular contributions to the total momentum.  

   As in sections \ref{ele:0} and \ref{mag:0}, we explain every step in detail.  This gives an opportunity to show how expressions containing both $\UPS$ and $\DUP$ functions can be integrated using simple techniques.

\subsection{Internal region $0 \leqslant  r \leqslant a$}

In the internal region $\vec{E}$ is the constant electric field \eqref{fin:8} and  $\vec{B}$ the constant magnetic field \eqref{fin:11}.  Thus, we have to calculate the vector product $\vec{E} \times \vec{B}$ where
\begin{eqnarray}
\label{app:2}
         \vec{E}_{dm}(\vec{r},r \leqslant a)
     &=  \vec{E}_0 \UPS(a-r)
       - \vec{u}(\vec{ E}_0 \cdot \vec{u})~ r \DUP(a-r),\\
\label{app:3}
         \vec{B}_{d}(\vec{r},r \leqslant b) 
     &=  \vec{B}_0 \UPS(b-r)
       - \dfrac{1}{2} \Bigl( \vec{B}_0 - \vec{u}
                   (\vec{B}_0 \cdot \vec{u})
                        \Bigr)~ r \DUP(b-r),
\end{eqnarray}
where we have used the identity $\vec{u}\times (\vec{B}_0 \times \vec{u}) = \vec{B}_0 - \vec{u}(\vec{B}_0 \cdot \vec{u})$ to replace the cross-products by a dot-product.  In the general case, this means that we have to integrate $6-1=5$ terms taking into account that $\vec{u} \times \vec{u} = 0$.  However, if we avoid the degenerate case $a=b$ and specialise to the case $b > a$, we remark that
\begin{eqnarray}\label{app:4}
     \UPS(b-r) = 1  \qquad\mbox{and}\qquad \DUP(b-r) = 0,
\end{eqnarray}
so that we need to integrate only two terms, namely
\begin{eqnarray}\label{app:5}
     \vec{E} \times \vec{B}
      = \vec{E}_0 \times \vec{B}_0 \UPS(a-r)
       -  \vec{u}(\vec{ E}_0 \cdot \vec{u})
           \times  \vec{B}_0~ r \DUP(a-r).
\end{eqnarray}
Integrating over angles, this is
\begin{eqnarray}\label{app:6}
    \frac{1}{4\pi} \iint d\omega~ \vec{E} \times \vec{B}
      = \vec{E}_0 \times \vec{B}_0 
        \Bigl( \UPS(a-r) - \frac{1}{3} r\DUP(a-r) \Bigr).
\end{eqnarray}
$\vec{P}_{\rm Poynting}$ is thus given by the radial integral
\begin{eqnarray}\label{app:7}
     \vec{P}_{\rm Poynting} = \vec{E}_0 \times \vec{B}_0 
    \int_0^{a-\alpha} dr \Bigl( r^2\UPS(a-r) - \frac{1}{3} r^3\DUP(a-r)\Bigr).
\end{eqnarray}
To integrate we notice that 
\begin{eqnarray}\label{app:8}
                   \Bigl(\frac{1}{3} r^3\UPS(a-r) \Bigr)'
        = r^2\UPS(a-r) - \frac{1}{3} r^3\DUP(a-r),
\end{eqnarray}
so that the radial integral becomes
\begin{eqnarray}\label{app:9}
     \vec{P}_{\rm Poynting} = \vec{E}_0 \times \vec{B}_0 
    \int_0^{a - \alpha} dr \Bigl(\frac{1}{3} r^3\UPS(a-r) \Bigr)',
\end{eqnarray}
which gives
\begin{eqnarray}\label{app:10}
     \vec{P}_{\rm Poynting} = \vec{E}_0 \times \vec{B}_0 
        \frac{1}{3} r^3\UPS(a-r) \Bigr|_0^{a - \alpha}.
\end{eqnarray}
Thus, using \eqref{fin:13}, i.e., $\vec{E}_0 = -  \vec{ p}/{a^3}$, we obtain finally
\begin{eqnarray}\label{app:11}
     \vec{P}_{\rm Poynting}(0 \leqslant  r \leqslant a-\alpha) = 
        \frac{1}{3} \frac{r^3}{a^3}\UPS(a-r) \Bigr|_0^{a - \alpha}
     \vec{B}_0 \times \vec{p}.
\end{eqnarray}

\subsection{Intermediate region $a \leqslant r \leqslant b$}

In the intermediate region $\vec{E}$ is the dipolar electric field \eqref{fin:2} and  $\vec{B}$ the constant magnetic field \eqref{fin:11}.  Thus, we have to calculate the vector product $\vec{E} \times \vec{B}$ where
\begin{eqnarray}
\label{app:12}
                 \vec{E}_{dm}(\vec{r},r \geqslant a)
     &= \Bigl( 3\frac{\vec{u}(\vec{p} \cdot \vec{u})}{r^3}
       - \frac{\vec{p}}{r^3} \Bigr)  \UPS(r-a)
   - \frac{\vec{u}(\vec{p}\cdot\vec{u})}{r^2} \DUP(r-a),\\
\label{app:13}
         \vec{B}_{d}(\vec{r},r \leqslant b) 
     &=  \vec{B}_0 \UPS(b-r)
       - \dfrac{1}{2} \Bigl( \vec{B}_0 - \vec{u}
                 (\vec{B}_0 \cdot \vec{u})
                        \Bigr)~ r \DUP(b-r),
\end{eqnarray}
where we expressed $\vec{B}_{d}$ as in \eqref{app:3}.  In the general case, this means that we have to integrate $9-2=7$ terms taking into account that $\vec{u} \times \vec{u} = 0$.  However, if we avoid the degenerate case $a=b$ we remark that
\begin{eqnarray}\label{app:14}
     \DUP(b-r)\DUP(r-a) = 0,
\end{eqnarray}
so that we need to integrate only six terms, namely
\begin{eqnarray}
\label{app:15}
                 \vec{E} \times \vec{B} = \frac{1}{r^3}\Bigl[~
               & 3\vec{u}(\vec{p} \cdot \vec{u})
                    \times \vec{B}_0 \UPS(r-a)\UPS(b-r)\\
\label{app:16}
               &- \frac{3}{2} \vec{u}(\vec{p}\cdot\vec{u})
                       \times \vec{B}_0  \UPS(r-a)r\DUP(b-r)\\
\label{app:17}
               &- \vec{p} \times \vec{B}_0  \UPS(r-a)\UPS(b-r)\\
\label{app:18}
               &+ \frac{1}{2} \vec{p} \times
                      \Bigl(\vec{B}_0 -\vec{u}
                           (\vec{B}_0 \cdot \vec{u}
                     )\Bigr) \UPS(r-a)r\DUP(b-r)\\
\label{app:19}
               &- \vec{u} (\vec{p}\cdot\vec{u})
                        \times \vec{B}_0 r\DUP(r-a)\UPS(b-r)~\Bigr].
\end{eqnarray}
Integrating over angles, this is
\begin{eqnarray}
\notag
         \frac{1}{4\pi} \iint d\omega \vec{E} \times \vec{B}
              = &- \frac{1}{r^2}\Bigl(
                   \frac{1}{6} \UPS(r-a)\DUP(b-r)\\
\label{app:20}
                &+ \frac{1}{3} \DUP(r-a)\UPS(b-r)
                   \Bigr)\vec{p} \times \vec{B}_0.
\end{eqnarray}
As $\UPS(r-a)\DUP(b-r)=\DUP(b-r)$ and $\DUP(r-a)\UPS(b-r)=\DUP(r-a)$ when $a\neq b$, we can remove the $\UPS$ functions. Thus, $\vec{P}_{\rm Poynting}$ is given by the radial integral
\begin{eqnarray}
\label{app:21}
    \vec{P}_{\rm Poynting} =  
   -\int_{a+\alpha}^{b-\beta} dr  \Bigl(
                \frac{1}{6} \DUP(b-r) + \frac{1}{3} \DUP(r-a)
                \Bigr)\vec{p} \times \vec{B}_0,
\end{eqnarray}   
which is simply
\begin{eqnarray}
\label{app:22}
     \fl \qquad \qquad
     \vec{P}_{\rm Poynting}(a+\alpha \leqslant r \leqslant b-\beta)
                = \Bigl(\frac{1}{3} \UPS(r-a)  - \frac{1}{6} \UPS(b-r)\Bigr)
                  \Bigr|_{a+\alpha}^{b-\beta}
                  \vec{B}_0 \times \vec{p}.
\end{eqnarray}

\subsection{External region $b \leqslant r \leqslant \infty$}

In the external region $\vec{E}$ is the dipolar electric field \eqref{fin:2} and  $\vec{B}$ the dipolar magnetic field \eqref{fin:5}.  Thus, we have to calculate the vector product $\vec{E} \times \vec{B}$ where
\begin{eqnarray}
\label{app:23}
     \fl \qquad \qquad
                 \vec{E}_{dm}(\vec{r},r \geqslant a)
      &= \Bigl( 3\frac{\vec{u}(\vec{p} \cdot \vec{u})}{r^3}
          - \frac{\vec{p}}{r^3} \Bigr)  \UPS(r-a)
  -  \frac{\vec{u}(\vec{p}\cdot\vec{u})}{r^2} \DUP(r-a),\\
\label{app:24}
     \fl \qquad \qquad
                 \vec{B}_{d}(\vec{r},r \geqslant b)
      &= \Bigl( 3\frac{\vec{u}(\vec{m} \cdot \vec{u})}{r^3}
          - \frac{\vec{m}}{r^3} \Bigr)  \UPS(r-b)
          + \frac{ \vec{m} - \vec{u}
                  (\vec{m} \cdot \vec{u}) }
                 {r^2} \DUP(r-b),
\end{eqnarray}
where we have used the identity $\vec{u}\times (\vec{m} \times \vec{u}) = \vec{m} - \vec{u}(\vec{m} \cdot \vec{u})$.  In the general case, this means that we have to integrate $12-4=8$ terms taking into account that $\vec{u} \times \vec{u} = 0$.   However, if we avoid the degenerate case $a=b$ and specialise to $b > a$, we remark that
\begin{eqnarray}\label{app:25}
     \UPS(r-a) = 1  \qquad\mbox{and}\qquad \DUP(r-a) = 0,
\end{eqnarray}
so we need to integrate only six terms, namely
\begin{eqnarray}
\label{app:26}
     \vec{E} \times \vec{B}
       &=  \frac{1}{r^6}\Bigl(
        - 3\vec{u}(\vec{p} \cdot \vec{u}) \times \vec{m}
        - 3\vec{p} \times \vec{u}(\vec{m} \cdot \vec{u})
        +  \vec{p} \times \vec{m}
                     \Bigr) \UPS(r-b)\\
\label{app:27}
      & +  \frac{1}{r^5}\Bigl(
        + 3\vec{u} \times \vec{m}(\vec{p} \cdot \vec{u})
        -  \vec{p} \times \vec{m}
        +  \vec{p} \times \vec{u}(\vec{m} \cdot \vec{u}
                      \Bigr) \DUP(r-b).
\end{eqnarray}
Integrating over angles, this is
\begin{eqnarray}\label{app:28}
    \frac{1}{4\pi} \iint d\omega \vec{E} \times \vec{B}
      = \vec{p} \times \vec{m} 
        \Bigl( -\frac{1}{r^6}\UPS(r-b) + \frac{1}{3r^5} \DUP(r-b) \Bigr).
\end{eqnarray}
$\vec{P}_{\rm Poynting}$ is thus given by the radial integral
\begin{eqnarray}\label{app:29}
     \vec{P}_{\rm Poynting} = \vec{p} \times \vec{m} 
  \int_{b+\beta}^\infty dr \Bigl(  -\frac{1}{r^4}\UPS(r-b)
                                     + \frac{1}{3r^3} \DUP(r-b) \Bigr),
\end{eqnarray}
which immediately integrates to
\begin{eqnarray}\label{app:30}
     \vec{P}_{\rm Poynting}
    = \vec{p} \times \vec{m} 
     \frac{1}{3r^3}\UPS(r-b)  \Bigr|_{b+\beta}^\infty.
\end{eqnarray}
Using \eqref{fin:14}, i.e., $\vec{m} = b^3 \vec{ B}_0/2$, we obtain finally
\begin{eqnarray}\label{app:31}
     \vec{P}_{\rm Poynting}(b+\beta \leqslant r \leqslant \infty) =
              \frac{1}{6} \frac{b^3}{r^3}\UPS(r-b)  \Bigr|_{b+\beta}^\infty
                            \vec{B}_0 \times \vec{p}.
\end{eqnarray}

\section*{References}

\section*{Errata to reference \cite{GSPON2004D}}

Four lines below equation (2.6) the erroneous expression $|r| = \sgn(r)$ should be replaced by $|r| = r~\sgn(r)$; in equation (3.4) the factors $d\theta d\theta$ should be replaced by $d\theta d\phi$; and the redundant minus signs in equations (4.4), (5.4), (5.5), (6.7) and (6.8) should be removed, so that the charge and current densities read as follows
\begin{eqnarray}
\nonumber
       4\pi \rho_m(\vec r) =    \vec \nabla \cdot \vec E_m 
                           =  e \frac{1}{r^2} \delta(r)
                           \qquad\qquad\qquad\qquad      &(4.4)\\
\nonumber
   4\pi \vec j_{d}(\vec r) = \vec\nabla \times \vec H_{d}(\vec r)
        =   3 \frac{\vec{\mu} \times \vec{r} }{r^4} \delta(r)
                           \qquad\qquad\qquad\qquad      &(5.5)\\
\nonumber
    4\pi \rho_{dm}(\vec r) = \vec\nabla \cdot \vec H_{dm}(\vec r)
        =   3 \frac{\vec{r} \cdot \vec{\mu} }{r^4} \delta(r)
                           \qquad\qquad\qquad\qquad      &(6.8)
\end{eqnarray}


\begin{thebibliography}{999}
\label{biblio}


\bibitem{JACKS2006-} Jackson J D 2006 Relation between interaction terms in electromagnetic momentum $\int d^3x \mathbf{E} \times \mathbf{B} /4\pi c$ and Maxwell's $e \mathbf{A}(x,t)/c$, and interaction terms of the field lagrangian $\mathcal{L}_{\rm em} = d^3x [\mathbf{E}^2 - \mathbf{B}^2 ]/8\pi$ and the particle interaction lagrangian, $\mathcal{L}_{\rm int} = e\phi - e \mathbf{v} \cdot \mathbf{A}/c$ \emph{Preprint} (8 May 2006) available at \verb!http://puhep1.princeton.edu/~mcdonald/examples/EM/jackson_050806.pdf! 


\bibitem{MCDON2006-} McDonald K T 2006 Electromagnetic momentum of a capacitor in a uniform magnetic field \emph{Preprint} (18 June 2006, revised 5 January 2007) available at\\
\verb!http://puhep1.princeton.edu/~mcdonald/examples/cap_momentum.pdf!

\bibitem{GSPON2004D} Gsponer A 2007 Distributions in spherical coordinates with applications to classical electrodynamics \emph{Eur. J. Phys.} {\bf 28} 267--275

\bibitem{FURRY1969-} Furry W H 1969 Examples of momentum distributions in the electromagnetic field and in matter \emph{Am. J. Phys.} {\bf 37} 621--636

\bibitem{HNIZD1997-} Hnizdo V 1997 Hidden momentum and the electromagnetic mass of a charge and current carrying body \emph{Am. J. Phys.} {\bf 55} 55--65

\bibitem{TANGH1962-} Tangherlini F R 1962 General relativistic approach to the Poincar\'e compensating stresses for the classical point electron \emph{Nuovo Cim.} {\bf 26} 497--524

\bibitem{JACKS1977-} Jackson J D 1977 On the nature of intrinsic magnetic dipole moments \emph{CERN report 77-17} (CERN, Geneva, 1 September 1977) 18 pp\\
   Jackson J D 1998 Reprinted in \emph{Physics and Society: Essays in Honour of Victor Frederick Weisskopf} ed V Stefan and V F Weisskopf (New York/Berlin: AIP/Springer) 236 pp

\bibitem{BARUT1993-} Barut A O and Xu B-W 1993 Derivation of the quantum Maxwell equations from relativistic particle Hamiltonian \emph{Int. J. Phys.} {\bf 32} 961--968

\bibitem{GSPON2006C} Gsponer A 2006 The locally-conserved current of the Li\'enard-Wiechert field \emph{Preprint} arXiv:physics/0612090 available at \verb!http://arXiv.org/pdf/physics/0612090!

\bibitem{JACKS1975-} Jackson J D 1975 Classical Electrodynamics 2nd  edn (New York: Wiley) 848 pp

\bibitem{ROMER1995-} Romer R H Electromagnetic field momentum \emph{Am. J. Phys.} {\bf 63} 777--778

\bibitem{GSPON2006B} Gsponer A 2006 A concise introduction to Colombeau generalised functions and their applications \emph{Preprint} arXiv:math-ph/0611069 available at \verb!http://arXiv.org/pdf/math-ph/0611069! 

\bibitem{AHARO1988-} Aharonov Y, Pearl P and Vaidman L 1988 Comment on ``Proposed Aharonov-Casher effect: Another example of an Aharonov-Bohm effect arising from a classical lag \emph{Phys. Rev. A.} {\bf 37} 4052--4055

\bibitem{HURNI2004-} Lanczos C 1919 The Relations of the Homogeneous Maxwell's Equations to the Theory of Functions --- A contribution to the theory of relativity and electrons (Doctoral dissertation, Typeseted by Hurni J-P with a preface by Gsponer A, 2004) \emph{Preprint} arXiv:physics/0408079 available at \verb!http://arXiv.org/pdf/physics/0408079!


\end{thebibliography}
\end{document}